\documentclass[10pt, a4paper, twocolumn, showabstract]{naverlabseurope}

\usepackage{lipsum}
\usepackage{multicol}
\usepackage{tikz,tkz-kiviat,pgfplots}

\usepackage[utf8]{inputenc} 
\usepackage[T1]{fontenc}    
\usepackage{hyperref}       
\usepackage{url}            
\usepackage{booktabs}       
\usepackage{amsfonts}       
\usepackage{nicefrac}       
\usepackage{microtype}      
\usepackage{xcolor}         
\usepackage{enumitem}[nolistsep]
\usepackage{multirow} 
\usepackage{array}
\usepackage[para]{footmisc}
\usepackage{bbm}
\usepackage{adjustbox}
\usepackage{pgfplots}
\usepackage{graphicx}
\usepackage{caption}
\usepackage{subcaption}
\usepackage{pgfkeys}
\usepackage[normalem]{ulem}
\usepackage{algpseudocode}
\usepackage{algorithm}
\usepackage{amsmath}

\newenvironment{customlegend}[1][]{%
    \begingroup
    \csname pgfplots@init@cleared@structures\endcsname
    \pgfplotsset{#1}%
}{%
    \csname pgfplots@createlegend\endcsname
    \endgroup
}%
\def\addlegendimage{\csname pgfplots@addlegendimage\endcsname}

\usepackage[show]{chato-notes}

\AtBeginDocument{%
  \providecommand\BibTeX{{%
    \normalfont B\kern-0.5em{\scshape i\kern-0.25em b}\kern-0.8em\TeX}}}

\newlist{inlinelist}{enumerate*}{1}
\setlist*[inlinelist,1]{label=\roman*),itemjoin={{, }},itemjoin*={{, and }}}

\makeatletter
\@namedef{ver@everyshi.sty}{}
\makeatother

\usepackage{pgfplots, pgfplotstable}
\pgfplotsset{compat=newest}
\usepgfplotslibrary{fillbetween}

\newcommand{\edc}{Blue}
\newcommand{\sdc}{Gray}
\newcommand{\effvc}{Goldenrod}
\newcommand{\bestc}{Black}
\newcommand{\redc}{Red}

\pgfmathsetmacro{\stdgrad}{30}

\tikzset{every mark/.append style={solid}}
\pgfplotsset{
	grid=both, width=\linewidth, try min ticks=5,
	legend cell align=left, legend style={fill opacity=0.8},
	ylabel near ticks,
    xlabel near ticks,
    every tick label/.append style={font=\footnotesize},
}

\pgfplotsset{
    edplot/.style={thick, color=\edc, mark=square*},
    sdplot/.style={thick, color=\sdc, mark=square*},
    bestplot/.style={thick, color=\bestc, mark=pentagon*},
    bestplot2/.style={thick, color=\bestc, mark=square*},
    effvplot/.style={thick, color=\effvc, mark=square*},
    fifthplot/.style={thick, color=\redc, mark=square*},
    pruningplot/.style={thick, color=\effvc, mark=pentagon},
    pruninglexicalplot/.style={thick, color=\sdc, mark=oplus},
    rerankpruninglexicalplot/.style={thick, color=\sdc, mark=oplus, dashed},
    rerankpruningplot/.style={thick, color=\effvc, mark=pentagon, dashed},   
    edrerankplot/.style={thick, color=\edc, mark=none, dashed},
    sdrerankplot/.style={thick, color=\sdc, mark=none, dashed},
    bestrerankplot/.style={thick, color=\bestc, mark=none, dashed},
    effvrerankplot/.style={thick, color=\effvc, mark=none, dashed},
    spladelexicalplot/.style={mark=square, only marks},
    spladeqlexicalplot/.style={mark=triangle, only marks},
    spladeplot/.style={mark=o, only marks},
    kcentplot/.style={mark=*, only marks, color=red, mark size=4pt},
    ktwofiveplot/.style={mark=oplus, only marks},
    k1kplot/.style={mark=x, only marks},
    }
\pdfoutput=1
\begin{abstract}
Learned sparse  models such as SPLADE have successfully shown how to incorporate the benefits of state-of-the-art neural information retrieval models into the classical inverted index data structure. Despite their improvements in effectiveness, learned sparse models are not as efficient as classical sparse model such as BM25. The problem has been investigated and addressed by recently developed strategies, such as guided traversal query processing and static pruning, with different degrees of success on in-domain and out-of-domain datasets.
In this work, we propose a new query processing strategy for SPLADE based on a two-step cascade. The first step uses a pruned and reweighted version of the SPLADE sparse vectors, and the second step uses the original SPLADE vectors to re-score a sample of documents retrieved in the first stage. Our extensive experiments, performed on 30 different in-domain and out-of-domain datasets, show that our proposed strategy is able to improve mean and tail response times over the original single-stage SPLADE processing by up to $30\times$ and $40\times$, respectively, for in-domain datasets, and by $12\times$ to $25\times$, for mean response on out-of-domain datasets, while not incurring in statistical significant difference in 60\% of datasets.
\end{abstract}

\begin{document}

\title{Two-Step SPLADE: Simple, Efficient and Effective Approximation of SPLADE}

\authors{Carlos Lassance$^{\star}$  \authsep
Hervé Dejean   \authsep
Stéphane Clinchant \authsep
Nicola Tonellotto$^{\dagger}$}

\affiliations{NAVER LABS Europe, $^{\dagger}$ University of Pisa}
\contributions{$^{\star}$ work done at Naver Labs, now at Cohere}
\website{https://github.com/naver}
\websiteref{\href{https://github.com/naver}}

%
%
%


\maketitle

\section{Introduction}
Learned Sparse Retrieval (LSR) models~\cite{deepimpactv1,nguyen2023unified,formal2022distillation,splade,sparta} aim at combining the best of two worlds: the traditional search infrastructure, based on an inverted index of interpretable terms, and the representation power of Pretrained Language Models (PLMs)~\cite{devlin2018bert}. Such models recompute term weights for documents and queries to improve effectiveness, going as far as learning how to expand texts to be even more effective in IR tasks. 
LSR underpinning hypothesis is that existing search infrastructure, namely the inverted index and its efficient algorithms can easily be used to serve such models~\cite{fntir2018}. However, a mismatch exists between the posting lists score distribution of LSR models and traditional models like BM25, leading to inefficiency issues~\cite{mackenzie2023efficient}. One could argue that first-stage retrieval for bag-of-word models has been heavily optimized for years leading to better algorithms such as MaxScore~\cite{turtle1995query}, WAND~\cite{broder2003efficient}, and Block-Max WAND~\cite{ding2011faster}; and that LSR models would need better optimisations when used with a traditional inverted index. Therefore, recent works based on Guided Traversal (GT) ~\cite{mallia2022faster,qiao2023optimizing} have adapted existing algorithms to improve the latency of LSR models, where the main goal is to reuse BM25 to guide the selection of scored documents during the retrieval phase.

In this work, we address the same research question: given the mismatch between LSR models and traditional search algorithms, how can we better use the existing search algorithms and how can we improve the latency of LSR models. While several learned sparse retrieval works have been proposed, we focus this work on SPLADE~\cite{splade}, due to its popularity and effectiveness. Our goal is to increase the flexibility of the end user (directly be able to modulate efficiency/effectiveness at retrieval time), while limiting the amount of change needed on the overall system for the provider of the search engine.

Following recent works, where LSR is used as first stage retriever in a multi-stage ranking pipeline~\cite{deepimpactv1}, the main goal of this work is to further split the first stage retrieval in two parts. More precisely, our method relies on the observation that SPLADE sparse vectors can be approximated by sparser vectors obtained with top pooling. Moreover, based on the discussion in~\cite{grand2020m} about saturation function and dynamic pruning, we add a term re-weighting to the SPLADE scoring function that improves the efficiency of dynamic pruning for SPLADE vectors via a trade-off with effectiveness\footnote{c.f. right part of Figure~\ref{fig:intersection}}. In other words, a very good approximation of a sparse vector is \textit{a sparser re-weighted version} of this vector. Therefore, we can first compute the approximated results with sparser vectors, extract a top $k$ sample and then only perform the full score computation with the \textit{original vectors} in this sample. Note that this is similar to what is done in an approximated nearest neighbor for dense retrieval, leading to an approximated nearest neighbor for sparse retrieval models. 

Overall, the contributions of this paper are the following:
\begin{itemize}
\item We show that SPLADE first-stage retrieval can be approximated by a two-step algorithm relying on sparser term re-weighted SPLADE vectors. 
\item Our approximation allows us to propose new ranking models as efficient as GT, i.e., between $12\times$ to $40\times$ faster than SPLADE but more effective than GT, i.e., with statistically significant gains in 50\% of the  tested datasets and without statistically significant losses in 87\% of the tested datasets.
 \end{itemize}
The remaining of this paper is organised as follows: Section 2 discusses the related work, Section 3 illustrates our Two-Step SPLADE, Section 4 presents our experiments, and Section 5 reports our conclusions.

\section{Related Works}
The re-usability hypothesis of LSR models, i.e., the hypothesis that learned sparse models may be seamlessly adopted with inverted indexes, has been a source of debate. In~\cite{mackenzie2023efficient}, the authors showed that while these new methods may be used in the current infrastructure, they are not at all optimized for it, with a large decrease in efficiency. This led to an influx of recent works~\cite{bruch2023approximate,lassance2022efficiency,mallia2022faster,qiao2023optimizing} that look into achieving cost/latency parity with BM25. Nevertheless, there often exists a performance drop or robustness issue to achieve that result. We separate these works by how they deal with dynamic pruning~\cite{fntir2018}.
In the first line of work (a), the LSR model is adapted to the current search algorithm, while, in the second case (b), the dynamic pruning/search implementation is adapted to the models. 

\paragraph{Adapted Models (a):} Efficient SPLADE~\cite{lassance2022efficiency}  propose a dedicated way to train more efficient models with L1 regularization on the query side and better pretraining while in LSR pruning~\cite{lassance2023static} standard pruning techniques are applied to LSR models and demonstrate efficiency gains. Even if such works demonstrate latency figures comparable to BM25, such methods still suffer from losses on out-of-domain data, i.e, zero shot on BEIR benchmark. We note that there exists a vast literature on static pruning that we could draw from~\cite{carmel2001static,buttcher2006document,altingovde2009practitioner,chen2013information} but we focus on the simpler methods from ~\cite{lassance2023static} that have already been shown to work well with SPLADE.

\paragraph{Adapted Search Mechanism(b):}
Term Impact Decomposition~\cite{mackenzie-etal-2022-accelerating} showed that splitting posting lists and completely redesigning retrieval leads to greatly improved efficiency without any effectiveness cost. Guided Traversal (GT) \cite{mallia2022faster} first showed how to use the BM25 scores to choose which documents to score, but it requires posting lists to be shared between BM25 and the LSR method, which excludes most LSR methods. This constraint is later relaxed in Optimized GT (OGT)~\cite{qiao2023optimizing}, which makes GT available to any LSR method, by changing from purely guidance as in GT to parametrized guidance, where weights are balanced between BM25 and LSR methods. OGT is further optimized for SPLADE in~\cite{qiao2023representation}, by adding a stricter pruning scheme during training based on soft and hard thresholds for documents and queries, which require specific model training on top of using OGT. 

\medskip
Beyond this dichotomy, other works aim to improve the effectiveness/efficiency trade-off. The GAR~\cite{macavaney2022adaptive} approach relies on graphs in document space and mixing the efficiency of sparse retrieval, e.g. BM25, and the effectiveness of cross-encoder rerankers, but still completely changes the architecture of retrieval and increases computational cost due to the use of cross-encoder rerankers. In the case of lexically-accelerated dense retrieval (LADR)~\cite{kulkarni2023lexically}, which combines GAR and GT by using BM25 as a first-stage for dense retrieval and are thus restricted by the choice of BM25 as the ``first-stage''. Finally, there are sketching and clustering works such as~\cite{bruch2023approximate,bruch2023bridging} which convert the sparse retriever into a dense retriever thanks to hashing, which is a completely different paradigm to the one we analyze here and would not let us continue to use the same architecture for retrieval.

\section{Two-Step SPLADE}
Given recent results~\cite{lassance2022efficiency,lassance2023static,qiao2023optimizing}, a prominent way to improve SPLADE latency has been to reduce the number of query terms, then by pruning document terms or by using GT. However, for very fast models ($\approx$ BM25 latency), there is actually a sharp drop in effectiveness, especially for out-of-domain experiments \footnote{cf experimental section}. Ideally, one would like to have at the same time i) fast retrieval; and ii) out-of-domain effectiveness.

To do so, we split our first-stage SPLADE ranking into two steps: \textbf{approximate} and \textbf{rescoring}. In the first step, both queries and documents are pruned or compressed aggressively to yield an efficient approximation to the original SPLADE model. Then, in the rescoring step\footnote{We prefer to use rescoring instead of reranking as it may confuse the reader with the cross-encoder reranking case.}, the initial top $k$ documents retrieved during the first step are rescored using the full documents and queries from the original SPLADE model.

\begin{figure*}
    \centering
    \includegraphics[width=1.0\textwidth]{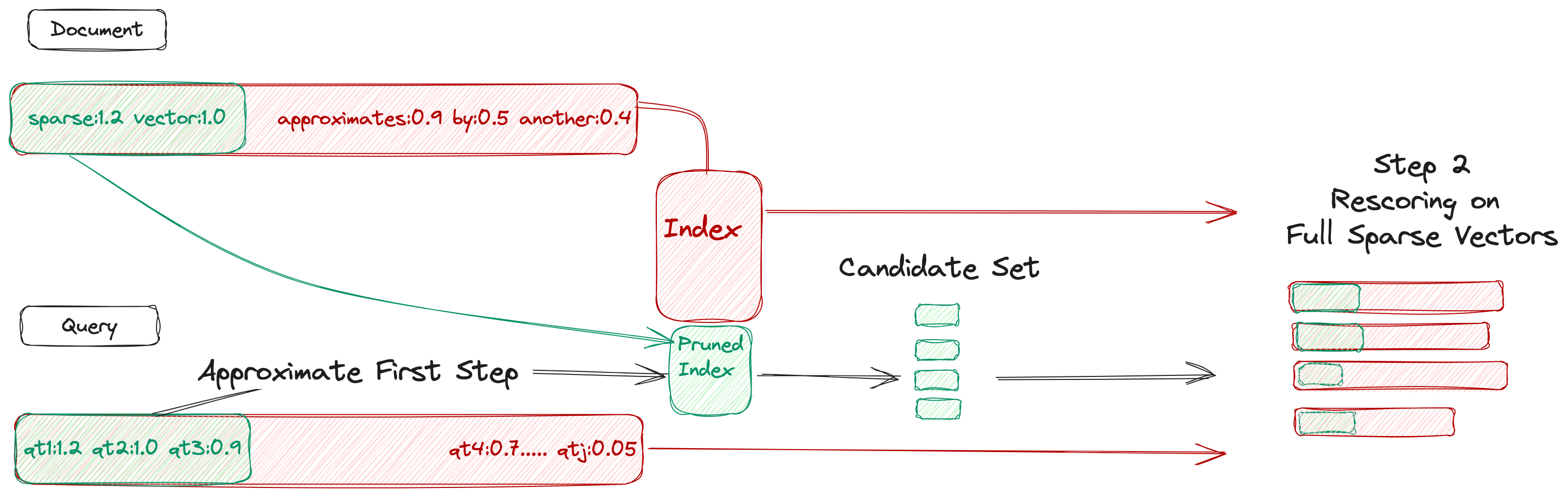}
    \caption{Two-Step SPLADE}
    \label{fig:twostage_splade}
\end{figure*}

This is indeed how most systems are implemented, with a ``first-stage`` and ''second-stage`` retrieval, where most of the time the first-stage is a vector-based retrieval and the second stage a cross-encoder. 
In other words, we actually propose to split the first-stage into two steps: the first step is a rough approximation of the representations (aggressive document and query pruning associated with term re-weighting) and the second step using the actual representations, although in a small subset of the original corpus. We illustrate the method in Figure \ref{fig:twostage_splade}.

\subsubsection{Indexing}

Starting from a trained SPLADE model, we use a top pooling as in~\cite{lassance2023static,shen2023lexmae,nguyen2023unified} to prune both documents and queries to the average size of the dataset (with upper limits of 128 and 32 tokens). For example, in the case of MSMARCO, we prune documents to the 50 tokens with the highest SPLADE-weights and queries to the highest 5. In other words, we rely on the observation that SPLADE sparse vectors can be approximated by sparser vectors based on the highest weights outputted by SPLADE. Note that hashing~\cite{bruch2023approximate} could in theory be used here to convert embeddings to a dense representation and perform classical dense retrieval ANN on it~\cite{bruch2023bridging} but that is outside the considerations of this work. Moreover, in GT~\cite{mallia2022faster,qiao2023optimizing}, i) the approximation step requires paired posting lists, including the same amount of query tokens; and ii) ranking is conducted with BM25 weighting, which is ideal for some datasets (especially ones where the annotation is based on BM25 results~\cite{thakur2021beir}), but not for all datasets, as we will see in our experiments. The indexing process is described in Algorithm~\ref{alg:pruning}

\begin{algorithm}
\caption{Indexing}\label{alg:pruning}
\begin{algorithmic}[1]
\Require{$T$ collection, $Q$ collection, preprocessed for SPLADE}
\State Compute averages of $T$ and $Q$ and store as $l_d$ and $l_q$
\State \textbf{Initialize} $I^a$ as an empty index // Approximate index
\State \textbf{Initialize} $I^r$ as an empty index // Rescoring index
\For{each document $d$ in $T$}
    \State $d' = $ Prune $d$ to size $l_d$ by selecting the highest values
    \State Add $d'$ to $I^a$ ; Add $d$ to $I^r$
    
\EndFor
\end{algorithmic}
\end{algorithm}

\subsubsection{Retrieval}

The retrieval process is divided into two steps: approximate and rescoring, and it is described in Algorithm~\ref{alg:query}.

\begin{algorithm}
\caption{Two-step Retrieval Algorithm}\label{alg:query}
\begin{algorithmic}[1]
\Require{Indexes $I^a$ and $I^r$, a query $q$, an average query size $l_q$, a saturation factor $k_1$ and the amount of documents to return $k$}
    \State $q' = $Prune $q$ to size $l_q$ by selecting the highest values
    \State $\textbf{scores}_{d'}$, $\textbf{D}$ = \textbf{SearchForTopkDocuments}($q', I^a, k, k_1$, null) // Approximate step:~Search on smaller index with smaller query, $k_1$ saturation, and without filters
    \State $\textbf{scores}_{d}$, $\textbf{D}$ = \textbf{SearchForTopkDocuments}($q, I^r, k, \infty, D$) // Rescoring step:~Search on larger index, without saturation, and with larger query while filtering to include only elements in $D$
    \State \textbf{Return} $\textbf{scores}_{d}$ and $\textbf{D}$
\end{algorithmic}
\end{algorithm}

\paragraph{Approximate step:}  we first rely on pruning, but also add a term-reweighting function. The main motivation for adding such a function is, as we will see, to get some speedup for the first step. Indeed, one of the main problems of LSR models like SPLADE is that the posting lists score distribution radically changes compared to BM25, making it harder for dynamic pruning to work effectively. This is due to the absence of the ``BM25’s saturation effect'' which is central, for example, to WAND~\cite{broder2003efficient,grand2020m}. 

In order to fix this we propose to reuse the BM25 term weighting, but with SPLADE-based term frequencies\footnote{The BM25 function would also have IDF, but we only consider the TF here.} in order to re-weight the SPLADE vectors. Note that this creates an additional approximation, as SPLADE itself is not trained for this scoring function. Considering that we have previously pruned the documents, the $b$ correction from BM25 is not necessary, leading to the following simpler formulation:
\begin{equation}
s(q,d) = \sum_{t \in q} \; B(t,q) \frac{\left(k_1+1 \right) TF\left(t,d \right)}{TF\left(t,d \right) + k_1}\;,
\end{equation}
where $s(q,d)$ is the score of document $d$ for query $q$, $t$ is a term in query $q$, $TF(t,d)$ is the term frequency/SPLADE weight of term $t$ on document $d$, $B(t,q)$ is the SPLADE weight of term $t$ on the query $q$, and $k_1$ is the saturation parameter. The higher the $k_1$ parameter, the more important the actual value of TF will actually be, while for smaller values of $k_1$, large values of TF will not be that important, achieving saturation via reduced difference of top TF values. Balancing $k_1$ means changing the scoring function from a constant, i.e. $k_1=0$, to heavily approximate SPLADE, i.e., $k_1=1$), to approximately the original SPLADE (as in the limit $k_1 \approx \infty$ the scoring simplifies to just the query and document term weights), i.e., $k_1=10,000$. Note that the more saturated our TFs are, the faster that retrieval will be when using dynamic pruning. We tested a large set of values and choose $k_1=100$ as the best trade-off, with $b$ set to 0.

\paragraph{Rescoring Step.}

In the second step, we rescore each document from the document set, selected by the first step, with the uncompressed, full sparse vectors (both query and document) coming from the initial SPLADE models. In other words, we are basically using the full SPLADE to rescore the top $k$ documents generated in the approximate step. In this work we always use $k=100$ (c.f. left part of Figure~\ref{fig:intersection}). Note that this rescorer can be easily used in current search engines such as PISA~\cite{mallia2019pisa} or Anserini~\cite{yang2017anserini}. In this work we implement restoring as a full index search that skips by document id using the `nextgeq` function from PISA, but many implementations are possible. This is different from a SPLADE GuidedTraversal that uses two different indexes for thresholding (BM25) and scoring (SPLADE).

\subsection*{Discussion}

The main benefit of our approach is that we can \textit{control} the quality or the speed of the approximation. Instead, GT is limited by the effectiveness of BM25 or requires the introduction of extra parameters to balance BM25 vs SPLADE~\cite{qiao2023optimizing}. We do not change anything on training time compared to~\cite{lassance2022efficiency,qiao2023representation}, meaning that our approach could be applied to any out-of-the-shelf LSR. We reuse pruning as suggested by~\cite{lassance2023static} but we also allow for control after the level of pruning is chosen (via term re-weighting and query pruning), without the need to index multiple pruned versions.

The main drawbacks of GT compared to our work are: 
i) the lack of flexibility of the approximation due to the use of BM25; ii) the need of posting list sharing between steps, making it complicated to use with BERT-based vocabularies and/or different terms between stages, e.g. a shortened query on the first-stage and full query on the second-stage.
Compared to GAR~\cite{macavaney2022adaptive} our approach has the advantage of not needing a cross-encoding reranker. However, we have the trade-off of indexing and searching on a SPLADE index (that we show is not slower than twice as BM25). Compared to sketching~\cite{bruch2023approximate,bruch2023bridging}, our approach has the advantage of staying in the sparse retrieval paradigm and not needing to change current infrastructure.

Finally, we discuss here some limitations of our study: 
a) while we reach almost parity with BM25 in terms of retrieval efficiency, we do not consider the cost of the query encoder, which would be closer to the dense retrievers from LADR~\cite{kulkarni2023lexically} and smaller than the cross encoding rerankers of GAR~\cite{macavaney2022adaptive}, but still large compared to retrieval cost; b) Our method also still requires some choice of parameters, such as the static and dynamic pruning strategies to employ.

Nonetheless we believe our method has the best trade-offs that we will showcase empirically in the next section. Ideally we desire the following properties: a) keep existing infrastructure intact; b) propose models with at most 2 times the retrieval latency as BM25; c) do not have a significant (t-test with p-value of 0.01) decrease of effectiveness compared to full SPLADE on at least half of datasets; d) allow for controllability of the trade-off at the end-user level.

\subsubsection*{Rescoring vs Reranking}

One thing that we do not touch in this work is the comparison of rescoring and reranking (with a cross-encoder or a stronger model). Both rescoring and reranking aim at improving the ranking quality via a trade-off of spending more computational power, they do so in different levels of cost. Rescoring only uses precomputed representations, meaning that it can be executed with low cost, while reranking does not normally allow for precomputed information. In other words, in a perfect world, one should be able to combine both (e.g rescoring top 1k and reranking top-10) and obtain scores that match costlier rerankings (e.g directly reranking the top 100 or 1000).

\subsubsection*{Storage requirements}

One drawback of the proposed two-step SPLADE implementation is that it requires storing two inverted-indexes: the full SPLADE index and the pruned SPLADE index. Compared to a SPLADE approach this is a small overhead as the pruned SPLADE indexes are in general much smaller than the full index, but it is still a nonzero difference. Compared to a SPLADE with Guided Traversal, the difference is almost negligible with the pruned SPLADE index taking around the same space as a BM25 index.

\subsubsection*{Effectiveness metric (Recall vs nDCG)}

In the following we do not report Recall values, only nDCG. We prefer to report nDCG for two reasons: 1) Easier to use across a multitude of datasets with different granularities of annotation; 2) A better recall does not mean that future reranking will be easier, especially with the current methods reranking methods that have similar biases to the first-stage.

\section{Experiments}
In our experiments, we want to demonstrate several findings:
\begin{enumerate}
    \item The validity of our approximation: we study several pruning and term re-weighting and show that our first-stage approximation is more accurate than BM25;
    \item Rescoring is a very efficient and effective technique to reduce the gap between the approximation and the full retrieval;
    \item Two-Step SPLADE (approximation + rescoring) is efficient and effective not only for in-domain experiments but also for maintaining its performance on zero-shot benchmarks (BEIR and LoTTe).
\end{enumerate}

\subsubsection{Search Engine and Dynamic Pruning}
 All of our experiments are done with the PISA~\cite{mallia2019pisa} search engine. PISA has different implementations of dynamic pruning algorithms and choosing which one to use is an important aspect of any system. We verify this by testing all our methods with the three most used algorithms: WAND~\cite{broder2003efficient}, Block-Max WAND~\cite{ding2011faster} and MaxScore~\cite{turtle1995query}. In the case of BM25 retrieval differences were small, with a slight advantage to MaxScore, while for original SPLADE (i.e. no pruning, no approximation), the best performing algorithm was also MaxScore which was twice as fast as the next one. However, pruning the SPLADE model, made Block-Max WAND was the most efficient. Finally, by adding term re-weighting on top of pruning, WAND becomes the most efficient. Due to the differences in these algorithms, we decided, in this work, to always report the result for the most efficient algorithm of the three. Code is available at \url{https://github.com/naver/splade/}

\subsubsection{Datasets}
We consider most datasets available in the literature to conduct our benchmark on effectiveness and efficiency. More importantly, we want to assess the out-of-domain effectiveness performance drop that most approximation methods incur.
In order to evaluate this, we compare methods on \textbf{30 datasets} \footnote{3 from MSMARCO, being the MSMARCO-dev, TREC-DL 19 and TREC-DL 20, 17 from BEIR~\cite{thakur2021beir} and 10 from LoTTe~\cite{santhanam2021colbertv2}}. Finally, note that for computing statistical significance we drop datasets that are composed of other datasets such as ``Pooled'' from LoTTe and CQADupStack from BEIR. Note that BEIR is heavily biased towards BM25, as noted on their own paper~\cite{thakur2021beir}, and thus we added LoTTe~\cite{santhanam2021colbertv2} exactly due to the fact that BM25 not being as effective. Ideally, a system should work on both.

Secondly, we use the Ranger library~\cite{sertkan2023ranger} to detect statistically significant changes ($p\leq 0.01$). In this case, we \textbf{count} the number of datasets for which a method is statistically significantly better, similar to a meta-analysis process~\cite{sertkan2023exploring}. 

\subsubsection{Models}
Finally, our study is based on SPLADE-v3 models\footnote{\url{https://huggingface.co/naver/splade-v3}} models as done in related works ~\cite{lassance2022efficiency,qiao2023optimizing}. In fact, the public SPLADE-v2 or SPLADE++ checkpoints are not state-of-the-art anymore. Therefore, we adopt the  SPLADE-v3, which are based on SPLADE++ Self-Distil but trained for slightly more time and with better distillation scores by adding L1 regularization as in \cite{lassance2022efficiency}.  

In terms of baselines, we compare our Two-Step SPLADE to a single-step pruned version, following~\cite{lassance2023static}, and a version of (Optimized) Guided Traversal~\cite{mallia2022faster,qiao2023optimizing}. Note that GT (our implementation) is not exactly GT, nor OGT, but something in the middle. Our implementation of GT is based on our SPLADE models and works in two separate phases: approximation BM25, and then rescoring full SPLADE.  We do so, because it would not be possible to have aligned posting lists between BM25 and our SPLADE (and thus traverse them concurrently) as in GT, and we did not want to modify the retrieval procedure as in OGT~\cite{qiao2023optimizing} which would require hyperparameter tuning for each dataset and using a different version of PISA compared to our other experiments. This baseline is a best-case scenario in terms of effectiveness for GT+SPLADE, while adding at most a 30\% penalty in efficiency when compared to BM25. 

\textbf{Additional comparisons:} EfficientSPLADE~\cite{lassance2022efficiency} is also matched against Two-Step SPLADE, most notably the available EFF-V-Large, but they are omitted from tables to avoid confusion between what SPLADE is being discussed. Finally, we do not compare directly to Term Impact Decomposition~\cite{mackenzie-etal-2022-accelerating} and Optimized Guided Traversal for SPLADE~\cite{qiao2023representation} because they require changes to the PISA software and SPLADE model, and while codebases are advertised in the respective papers, they were not available at the time of writing. However, we do present comparison numbers against both OGT~\cite{qiao2023optimizing,qiao2023representation}, but only as a hint of relative efficiency/effectiveness, as the comparison is not totally fair due to not using the same machine/setting.

\subsection{Approximation Validity}

We study the validity of the approximation step for SPLADE. To do so, we measure the top $k$ intersection of the approximate step with the original top $k$ from SPLADE. First, choosing the optimal pruning values turns out to be complicated when dealing with a multitude of datasets. Therefore, we will assess our use of a simple heuristic: the original statistics from the dataset. Secondly, for the approximation step we can vary the term re-weighting function in order to play with the efficiency/effectiveness trade-off.

\subsubsection{Approximate Step Parameters}

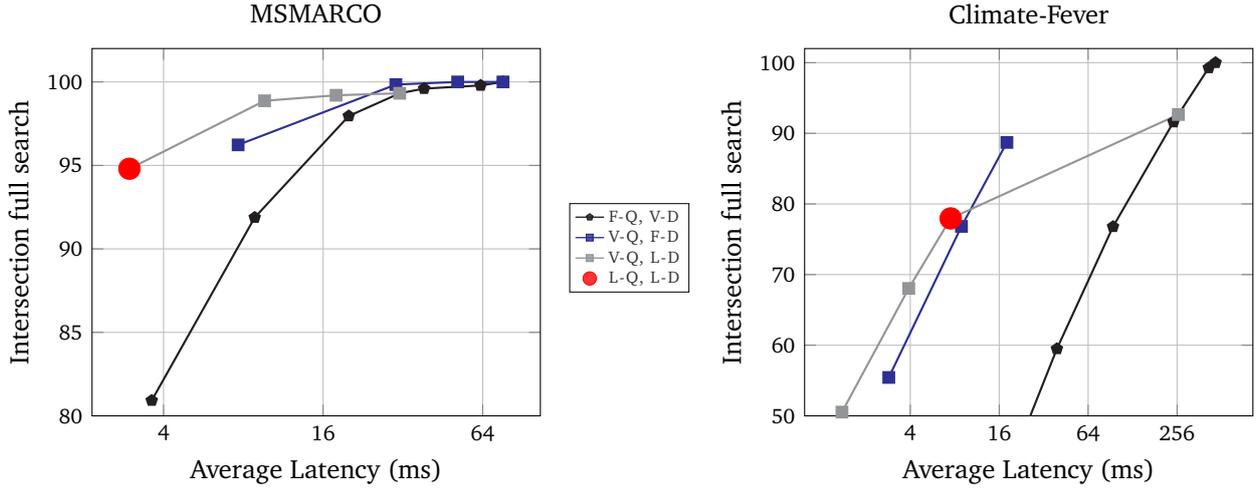
\begin{figure*}[ht]
     \centering
     \begin{subfigure}[t]{0.44\textwidth}
         \centering
\adjustbox{max width=\textwidth}{%
            \begin{tikzpicture}
       \begin{axis}[
           xlabel={Average Latency (ms)},
           ylabel=Intersection full search,
           title=MSMARCO,
           ymin=80.0, ymax=102.0,
            xmode=log,
            log ticks with fixed point,log basis x={2},
            xtick={4,16,64},
                      xticklabels = {4,16,64}           
           ]
         \addplot[bestplot] table {tikz/intersection/pruning/m-fq-vd.txt};
         \addplot[edplot] table {tikz/intersection/pruning/m-vq-fd.txt};
         \addplot[sdplot] table {tikz/intersection/pruning/m-vq-ld.txt};
         \addplot[kcentplot] table {tikz/intersection/pruning/m-lq-ld.txt};
       \end{axis}
    \end{tikzpicture}
         }
     \end{subfigure}
     \begin{subfigure}[t]{0.1\textwidth}
         \centering
\adjustbox{max width=\textwidth}{%
         \raisebox{120px}{%
         \begin{tikzpicture}
\begin{customlegend}[
legend columns=1,
legend entries={
\textsc{F-Q, V-D},
\textsc{V-Q, F-D},
\textsc{V-Q, L-D},
\textsc{L-Q, L-D}
                        }]
         \addlegendimage{bestplot};
         \addlegendimage{edplot};
         \addlegendimage{sdplot};
         \addlegendimage{kcentplot};
        \end{customlegend}
\end{tikzpicture}}
}
     \end{subfigure}
     \begin{subfigure}[t]{0.44\textwidth}
         \centering
\adjustbox{max width=\textwidth}{%
            \begin{tikzpicture}
       \begin{axis}[
           xlabel={Average Latency (ms)},
           ylabel=Intersection full search,
           title=Climate-Fever,
           ymin=50.0, ymax=102.0,
            xmode=log,
            log ticks with fixed point,log basis x={2},
            xtick={4,16,64,256},
                      xticklabels = {4,16,64,256}           
           ]
         \addplot[bestplot] table {tikz/intersection/pruning/c-fq-vd.txt};
         \addplot[edplot] table {tikz/intersection/pruning/c-vq-fd.txt};
         \addplot[sdplot] table {tikz/intersection/pruning/c-vq-ld.txt};
         \addplot[kcentplot] table {tikz/intersection/pruning/c-lq-ld.txt};
       \end{axis}
    \end{tikzpicture}
         }
     \end{subfigure}

      \caption{Intersection of the original retrieval top-10 and the pruned retrieval. We represent the different combinations by: F: No pruning; V: Varying pruning; L: Lexically-pruned. D: Document, Q: Query. So F-Q, V-D means we vary document length and keep the query untouched.}
     \label{fig:varying-pruning}
\end{figure*}

During indexing,  we can now control the level of pruning for the approximate stage. Ideally, we would want something that already speeds up retrieval substantially, without losing effectiveness. Following~\cite{lassance2023static} this would be around half or a quarter of the original SPLADE document size. In order to check this we look into SPLADE retrieval of MSMARCO-dev and climate-FEVER, and prune documents (V-D) into multiple sizes ($8$,$16$,$32$,$64$,$128$,$N/A$) and also to the lexical size (L-D) of the dataset (50 for MSMARCO and 67 for climate-fever). We also include query pruning (V-Q) with different values ($5$,$10$,$16$,$N/A$), with 5 and 16 being the lexical size (L-Q) for MSMARCO and Climate-FEVER. Note that such query pruning was not included in the previous work. Results are presented in Figure~\ref{fig:varying-pruning}.

What we first notice is that the effects are very different depending on the dataset. MSMARCO seems quite easy to prune, with our methods easily keeping more than 90\% intersection of the top 10 (i.e. in average 9 out of the top10 documents from the original search are in the top $100$ of the approximate search), however, the results are not as good in climate-FEVER. In other words if we selected the values solely using the characteristics of MSMARCO the method would underperform in climate-fever and vice-versa. This effect is what motivates us to test in so many datasets and to look into both efficiency and effectiveness in all of them and not solely in MSMARCO (as many methods do). Nonetheless, a simple heuristic that seemed good enough to us is just \textit{using the average token size} (document and query), which we call lexical size (red dot).

\subsubsection{Fixed Pruning, change $k_1$}

\begin{figure*}[ht]
     \centering
     \begin{subfigure}[t]{0.39\textwidth}
         \centering
\adjustbox{max width=\textwidth}{%
            \begin{tikzpicture}
       \begin{axis}[
           xlabel={Rescored top-$k$},
           ylabel=Intersection full search,
           title=Varying top-$k$,
           ytick={40,50,60,70,80,90,100},
           yticklabels={40,50,60,70,80,90,100}, ]
         \addplot[bestplot2] table {tikz/intersection/k1/top-k-inf.txt};
         \addplot[edplot] table {tikz/intersection/k1/top-k-400.txt};
         \addplot[sdplot] table {tikz/intersection/k1/top-k-100.txt};
         \addplot[effvplot] table {tikz/intersection/k1/top-k-10.txt};
         \addplot[fifthplot] table {tikz/intersection/k1/top-k-bm25.txt};
       \end{axis}
    \end{tikzpicture}
         }
     \end{subfigure}
     \begin{subfigure}[t]{0.2\textwidth}
         \centering
\adjustbox{max width=\textwidth}{%
         \raisebox{80px}{%
         \begin{tikzpicture}
\begin{customlegend}[
legend columns=1,
legend entries={
\textsc{P-SPLADE,$k_1=\infty$},
\textsc{P-SPLADE,$k_1=400$},
\textsc{P-SPLADE,$k_1=100$},
\textsc{P-SPLADE,$k_1=10$},
\textsc{BM25}
                        }]
         \addlegendimage{bestplot2};
         \addlegendimage{edplot};
         \addlegendimage{sdplot};
         \addlegendimage{effvplot};
         \addlegendimage{fifthplot};
        \end{customlegend}
\end{tikzpicture}}
}
     \end{subfigure}
     \begin{subfigure}[t]{0.39\textwidth}
         \centering
\adjustbox{max width=\textwidth}{%
            \begin{tikzpicture}
       \begin{axis}[
           xlabel={AvG L from BEIR > 1M},
           ylabel=Intersection full search,
           title=Rescore top-$100$,
           ]
    \addplot[
    scatter/classes={a={\bestc}, b={\edc}, c={\sdc}, d={\effvc}, e={\redc}},
    scatter,
    mark=square*,
    only marks,%
    visualization depends on=\thisrow{ey} \as \myshift,
    every node near coord/.append style = {shift={(axis direction
    cs:0,\myshift)}},
    scatter src=explicit symbolic,
    visualization depends on={value \thisrow{label} \as \Label},
    ]%
    plot [error bars/.cd, y dir = both, y explicit]
    table[meta=class, x=x, y=y, y error=ey]{
        x   y   ey  class   label
        15.492756 94.12942923132614 2.481509576379759   a   $k_1=\infty$
        11.279194 93.65746805592056 2.4282948551261785   b   $k_1=400$
        7.08164 91.06207156225535 2.7102699651466833   c   $k_1=100$
        4.406679 84.72339754853007 3.441369983545317   d   $k_1=10$
        4.275297 67.06141382172677 4.487817134579153   e bm25
    };
       \end{axis}
    \end{tikzpicture}
         }
     \end{subfigure}

      \caption{Intersection of the original retrieval top-10 and the first approximate step. P-SPLADE means the L-Q, L-D version from the previous figure with the applied $k_1$ saturation. AvG L is the average latency for BEIR datasets that have more than 1M documents.}
     \label{fig:intersection}
\end{figure*}
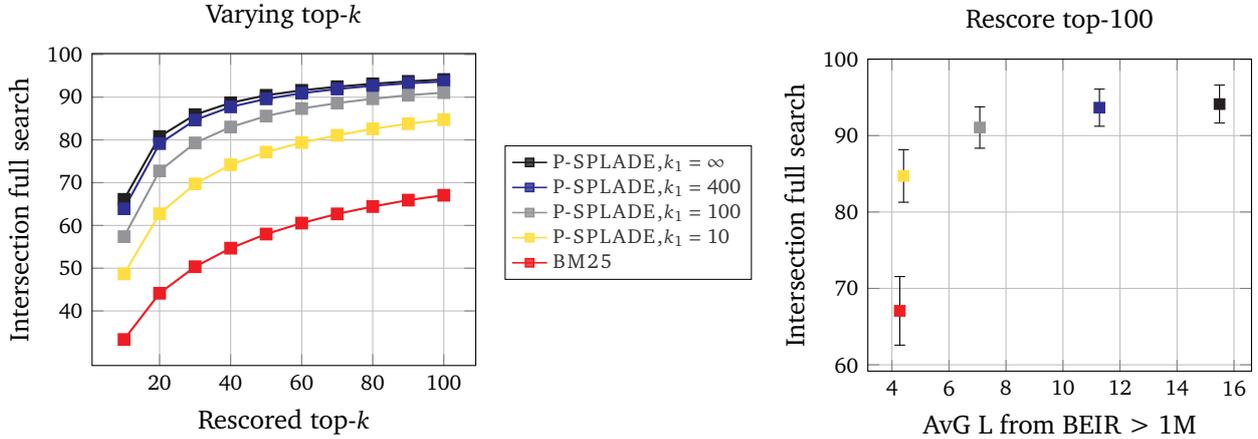

Now that we have chosen the pruning, we can use all our datasets to test different versions of term re-weighting functions by varying $k_1$. We present the results in Figure~\ref{fig:intersection}, where on the left we look into varying the top $k$ retrieval and $k_1$ values and on the right we fix $k=100$ and display the efficiency-effectiveness trade-off of different values of $k_1$, with the efficiency being measured as the average of average latencies over large BEIR datasets (>1M documents). The results show that the larger $k_1$ is, the more accurate the approximate is, but the larger $k_1$ is, the larger the latency is. Therefore, the saturation function allows to easily control the efficiency-effectiveness trade-off. Finally, a top $k$ of 100 and a saturation with $k_1=100$ seem to be appropriate parameters, as they achieve around 91\% intersection with the ``full-scale'' retrieval, with a .99 confidence interval between 88\% and 94\% intersection, while keeping efficiency close to BM25/GT.

\subsection{Effectiveness and Efficiency Benchmark}
After having demonstrated the validity of our approximation, and analysed the parameter's sensitivity, we can now proceed with the full benchmark. We present results by \textit{statistical differences and relative latency to the BM25 baseline} in Table~\ref{tab:summary}, while raw latencies and average effectivenesses are presented in Table~\ref{tab:summary_all}. From these tables of results, we draw the following conclusions:

\begin{enumerate}
    \item Approximate SPLADE (row e): term re-weighting improves efficiency in out-of-domain datasets compared to the pruned-only version (row c) (1.1 vs 2.3 and 1.4 vs 2.7 on BEIR and LoTTe, but reduces effectiveness.
    \item Two Step SPLADE (row f and g):  can keep reasonable retrieval efficiency compared to BM25 (less than 2 times average latency increase with the approximate SPLADE) while having similar effectiveness to the full SPLADE (row b). Note that this means a latency improvement of 12x to 40x on full SPLADE depending on the dataset and that in MSMARCO it is even faster than BM25, especially when considering p99 latency.
    \item Comparison to previous work (row (c)  and (d) GT ): we see clear improvements in effectiveness and efficiency. Against GT, TwoStage is strictly superior for 14 out of 30 datasets and only worse on 4, while having similar latencies.
\end{enumerate}

\begin{table*}[ht]
\centering
\caption{Effect size analysis ($p\leq 0.01$) of methods against SPLADE (b) and our implementation of Guided Traversal (GT) (d). $\geq$ is the number of datasets where the row model does not present statistical drop of effectiveness against the column one, > the ones with statistical improvement and < statistical drop. Latency is normalized by BM25 (1.0 is $=$ to BM25) and thus is the lower the better. >1M means the subset of datasets of BEIR that have at least 1M documents on the corpus. AvG L is the average latency of the dataset or the average of dataset averages (for BEIR and LoTTe). }
\label{tab:summary}

\begin{tabular}{@{}cc|cc|cc|cc|cc|c@{}}
\toprule
\multirow{3}{*}{} & \multirow{3}{*}{Method} & \multicolumn{4}{c|}{Effect   size against}             & \multicolumn{2}{c|}{MSMARCO}                        & \multicolumn{2}{c|}{BEIR} & \multirow{2}{*}{Lotte} \\ \cmidrule(l){3-10} 
                  &                         & \multicolumn{2}{c|}{SPLADE (b)} & \multicolumn{2}{c|}{GT (d)} & \multicolumn{2}{c|}{Latency} & 18    & \textgreater{}1M &       \\
                  &                         & $\geq$ ($>$)    & $<$  & $\geq$ ($>$)    & $<$    & Average   & p99                      & \multicolumn{2}{c|}{AvG L}              & AvG L         \\ \midrule
\multicolumn{11}{c}{Baselines}   \\
\midrule 
a & BM25                & 7 (1)  & 23 & 7 (1) & 23   & 1.0  & 1.0  & 1.0  & 1.0  & 1.0   \\
b & SPLADE-v3       &       \multicolumn{2}{c|}{N/A} & 27 (16)  & 3    & 19.1 & 12.4 & 24.8 & 32.6 & 22.1 \\
\midrule 
\multicolumn{11}{c}{Advanced Baselines}   \\ \midrule 
c & Approx. First Step \cite{lassance2023static} over (b)      & 7 (1) & 23 & 16 (2) & 14  & 0.7  & 0.4  & 2.3  & 2.6  & 2.6   \\
d & GT (Our Implementation) ($a \xrightarrow{} b$) & 14 (3) & 16  & \multicolumn{2}{c|}{N/A} & 1.1  & 1.0  & 1.2  & 1.2  & 1.3 \\
\midrule 
\multicolumn{11}{c}{This work}   \\
\midrule 
e & Approx. First Step (c) with $k_1=100$           & 4 (1)  & 26 & 5 (1) & 25  & 0.5 & 0.3 & 1.1 & 1.2 & 1.6 \\
f & Two-Step ($c \xrightarrow{} b$)      & 22 (0) & 8 & 26 (15)  & 4 & 0.8 & 0.4 & 2.5 & 2.7 & 3.0 \\
g & Two-Step ($e \xrightarrow{} b$) & 18 (2) & 12 & 26 (14)  & 4  & 0.6 & 0.3 & 1.4 & 1.3 & 1.8 \\
\bottomrule
\end{tabular}
\end{table*}

Concerning latencies, on the table we report p99 solely to MSMARCO as averaging p99 does not make sense. For p99 over BEIR our method (row g) is faster than BM25 (row c) in 9 out of the 17 datasets (excluding CQAdupstack) and 3 out of the 10 in LoTTe. We also separate the average latencies into all and just the largest in BEIR to isolate the smaller and larger dataset contributions and show that our method is stable over dataset sizes.

It is also interesting to look into the raw numbers in Table~\ref{tab:summary_all}. If we only looked into average numbers, and not the statistical differences over a large collection of datasets, conclusions would change. For example, in average terms, there is almost no difference on BEIR between SPLADE, GT and Two-Step. However, when we add LoTTe to the mix and look into it dataset by dataset, and not as an average of nDCG@10 as per the current usage in most papers, we can see the differences between methods\footnote{Due to lack of space, we had to omit the figure showing per-dataset comparison.}.

\begin{table*}[ht]
\centering
\caption{Efficiency and Effectiveness Benchmark: represented by average statistics and latency in ms instead of reporting relative to a baseline as in Table 1.}
\label{tab:summary_all}
\begin{tabular}{@{}cc|cc|cc|cc@{}}
\toprule
\multirow{3}{*}{} & \multirow{3}{*}{Method} & \multicolumn{2}{c}{MSMARCO}                                  & \multicolumn{2}{c}{BEIR}        & \multicolumn{2}{c}{Lotte}                             \\ \cmidrule(l){3-8}
                    &                         & \multicolumn{2}{c|}{Dev   Set}   & \multicolumn{2}{c|}{18 datasets} & \multirow{2}{*}{AvG L}        & \multirow{2}{*}{Success@5}\\
                    &                         & AvG L & MRR@10 & AvG L            & nDCG@10        &      \\ \midrule
\multicolumn{8}{c}{Baselines}                                                                                                                                                                          \\ \midrule
a & BM25                & 4.0  & 18.7 & 3.2  & 41.4 & 1.4  & 52.9 \\
b & SPLADE-v3             & 76.3 & 40.2 & 78.5 & 48.4 & 31.0 & 70.4  \\
\midrule 
\multicolumn{8}{c}{Advanced Baselines}   \\ \midrule 
c & Approx. First Step \cite{lassance2023static} over (b)     & 3.0  & 37.2 & 7.4  & 45.1 & 3.7  & 66.3  \\
d & GT (Our Implementation) ($a \xrightarrow{} b$)  & 4.4   & 36.7 & 3.5  & 48.2 & 1.8  & 66.4  \\ \midrule

\multicolumn{8}{c}{This work}                                                                                                                                                                          \\ \midrule
e & Approx. First Step (c) with $k_1=100$           & 2.0 & 35.2 & 3.6 & 42.2 & 2.2 & 63.1  \\
f & Two-Step ($c \xrightarrow{} b$)       & 3.3 & 40.1 & 8.1 & 48.0 & 4.2 & 70.1  \\
g & Two-Step ($e \xrightarrow{} b$) & 2.4 & 40.0 & 4.3 & 47.6 & 2.5 & 70.0  \\ \bottomrule
\end{tabular}
\end{table*}

\subsubsection{Additional Comparisons}
First, we compare our model to EFF-V-Large~\cite{lassance2022efficiency} (not shown on tables for clarity), where our method is faster (around 8x on MSMARCO) and more effective (0 datasets with statistical significant drop and 22 improvements). Furthermore, we note that we have also done all the previous sets of experiments using a SPLADE that is publically available: SPLADE++ CoCondenser Self-Distil from ~\cite{formal2022distillation} and have noted similar results (same effectiveness in 17 out of 30 instead of 18), although with higher latencies due to the different statistics of the models. We omit these tables due to lack of space.

Finally, looking into raw values we can be tempted to compare with raw numbers from other papers, such as~\cite{qiao2023optimizing,qiao2023representation}. These papers report a MSMARCO latency of 22 and 6.9ms for retrieving the top $10$ documents, while we report 2.4 ms with our method to retrieve the top $100$ documents. Looking at zero-shot over BEIR, we can compare the speed-up over the baseline, where our method improves more than 15 times over running the full SPLADE, while~\cite{qiao2023optimizing} improves it by 2.7 times and ~\cite{qiao2023representation} improves it by 3.6 times. However, due to experiments being run on different machines and base models, these comparisons only hint that our model is better, but cannot be concluded without proper experiments.

\section{Conclusion}

In this work, we have shown that by separating the first-stage of SPLADE into two steps we are able to mostly conserve its effectiveness and reduce latency by 12x to 40x. The proposed approach is more flexible than existing approaches while being directly applied to any search engine as a Two-Step search (unfiltered and then filtered to the top $k$ of the first search). This is another step into making LSR models more production-ready, which seems to show that in terms of retrieval latency LSR methods can compete with BM25 without problem. There are however two last barriers due to document/query inference i) indexing time and ii) actual full-scale latency measurements including query inference time, that we do not touch in this work and leave as future work.

\section*{Acknowledgements}
This work is supported, in part, by the spoke ``FutureHPC \& BigData'' of the ICSC – Centro Nazionale di Ricerca in High-Performance Computing, Big Data and Quantum Computing, the Spoke ``Human-centered AI'' of the M4C2 - Investimento 1.3, Partenariato Esteso PE00000013 - "FAIR - Future Artificial Intelligence Research", funded by European Union – NextGenerationEU, the FoReLab project (Departments of Excellence), and the NEREO PRIN project funded by the Italian Ministry of Education and Research Grant no. 2022AEFHAZ.

{
    \small
    \bibliographystyle{ieeenat_fullname}
    \bibliography{sample-base}
}


\appendix

\end{document}